\documentclass{svjour3}

\usepackage[english]{babel}
\usepackage{amssymb}
\usepackage{graphicx}
\usepackage{lineno}
\usepackage{amsmath}

\smartqed

\begin{document}

\title{Multi-lump solutions of mKP-1,2 equations with integrable boundary condition via $\overline{\partial}$-dressing}
\titlerunning{Multi-lump solutions of mKP-1,2 equations with integrable boundary condition}
\author{V.G. Dubrovsky \and A.V. Topovsky}
\institute{V.G. Dubrovsky \at
              Novosibirsk State Technical University, Karl Marx prospect 20, 630072, Novosibirsk, Russia. \\
              \email{dubrovskij@corp.nstu.ru}
           \and
          A.V. Topovsky \at
          Novosibirsk State Technical University, Karl Marx prospect 20, 630072, Novosibirsk, Russia. \\
          \email{topovskij@corp.nstu.ru}     }
\date{Received: date / Accepted: date}
\maketitle

\begin{abstract}
We constructed new classes of exact multi-lump solutions of mKP-1,2 equations with integrable boundary condition
$u(x,y,t)\big|_{y=0}=0$ by the use of $\overline\partial$-dressing method of Zakharov and Manakov. We exactly satisfied  reality and boundary conditions for the field $u(x,y,t)$
using general determinant formula for multi-lump solutions.
We illustrated  new calculated classes  by simple examples of two-lump solutions and demonstrated how fulfilment of integrable boundary condition $u\big|_{y=0}=0$ via special nonlinear superposition of several single lumps leads to formation of certain eigenmodes for the field $u(x,y,t)$ in semiplane $y\geq0$,  the analogs of standing waves on the string arising from corresponding boundary conditions at endpoints of string.

\end{abstract}

\section{Introduction}
\label{Section_1}
\setcounter{equation}{0}
The mKP-equation will be used in present paper in the following form~\cite{Konopelchenko_mKp}, \cite{JimboMiwa}:
\begin{equation}\label{mKP}
u_{t}+ u_{xxx}-3\sigma^2\left(\frac{1}{2}u^2u_x-\partial_x^{-1}u_{yy}+u_x\partial_x^{-1}u_{y}\right)=0,
\end{equation}
with $\sigma=i$ to mKP-1 and $\sigma=1$ to mKP-2 versions. The mKP equation (\ref{mKP}) can be represented as compatibility condition in well known Lax form $\left[L_1,L_2\right]=0$ of two linear auxiliary problems~\cite{Konopelchenko&Dubrovsky}, \cite{Konopelchenko&Dubrovsky2}:
\begin{equation}\label{mKP auxiltary problems}
\left\{
\begin{array}{ll}
L_1\psi=\sigma\psi_y+\psi_{xx}+V\psi_x=0, \\
L_2\psi=\psi_t+4\psi_{xxx}+6V\psi_{xx}+3\left(V_x-\sigma\partial^{-1}_x V_y+\frac{1}{2}V^2\right)\psi_x+\alpha\psi=0,
\end{array}
\right.
\end{equation}
here $V=\sigma u$. Several classes of exact solutions for both versions of mKP equation have been constructed at first in the paper of Konopelchenko and Dubrovsky~\cite{Konopelchenko&Dubrovsky}, \cite{Konopelchenko&Dubrovsky2} by the use of local and nonlocal Riemann-Hilbert problems \cite{Manakov}-\cite{Zakharov&Manakov} and more general $\overline{\partial}$-dressing method of Zakharov and Manakov \cite{Beals&Coifman1}-\cite{KonopelchenkoBook2}: with functional parameters, multi-solitons and multi-lumps.

The theme of construction of exact solutions with boundary conditions in broad sense: for linear, nonlinear equations, integrable or not, very actual in mathematical and theoretical physics. The concept of integrable boundary conditions for integrable nonlinear equations compatible with integrability of these equations  was at first introduced by Sklyanin~\cite{Sklyanin}. In subsequent papers of Habibulin et al ~\cite{Habibullin}-\cite{HabibullinKDV} and others~\cite{Vereshchagin} the concept of integrable boundary conditions to several types of integrable nonlinear equations has been applied: for difference equations, (1+1)-dimensional and (2+1)-dimensional  nonlinear differential
and integro-differential equations. A list of integrable boundary conditions for known (2+1)-dimensional integrable  KP, mKP, Veselov-Novikov, Ishimori, etc. equations has been proposed and examples of exact solutions for these integrable nonlinear equations  have been calculated ~\cite{Habibullin}-\cite{Vereshchagin}.
A. S. Fokas and his collaborators \cite{FokasBook} interesting results have been obtained via so called Unified Approach to Boundary Value Problems for one-dimensional and multi-dimensional linear and nonlinear differential equations.
The effectiveness of $\overline\partial$-dressing method of Zakharov and Manakov for construction of exact solutions of 2+1-dimensional integrable nonlinear equations with integrable boundaries was at first demonstrated in \cite{Dubrovsky&Topovsky&Ostreinov}, there  new exact multi-solitons and periodical solutions of KP-2 equation with integrable boundary condition $(u_{xx}(x,y,t)+\sigma u_y(x,y,t))|_{y=0}=0$ have been calculated.

In the present paper we constructed new classes of exact multi-lump solutions of mKP equation (\ref{mKP}) with integrable boundary condition~\cite{Habibullin2}
\begin{equation}\label{BoundaryCond}
u(x,y,t)\big|_{y=0}=0
\end{equation}
 via $\overline\partial$-dressing method of Zakharov and Manakov~\cite{Zakharov}-\cite{KonopelchenkoBook2}.
We derived general determinant formula for these solutions and using it satisfied the condition of reality $u=\overline u$ and boundary condition
(\ref{BoundaryCond}) in explicit form.

The paper is organized as follows. The first section is introduction. In second section we reviewed the basic formulae of $\overline\partial$-dressing for mKP equation (\ref{mKP}) and derived general determinant formula in convenient form for calculations of multi-lump solutions for both versions of  mKP equation. In third and fourth sections we obtained the restrictions on parameters of complex-valued exact solutions from boundary and reality conditions using corresponding determinant formulas.  In fifth and sixth sections we constructed new classes of real multi-lump solutions with integrable boundary for mKP-1 and mKP-2 versions of mKP equation. Futhermore we illustrated  new calculated classes  by simple examples of two-lump solutions. We demonstrated that satisfaction to integrable boundary condition $u\big|_{y=0}=0$ via special nonlinear superposition of several single lumps leads to formation of certain eigenmodes for the field $u(x,y,t)$ in semiplane $y\geq0$,  the analogs of standing waves on the string arising from corresponding boundary conditions at endpoints of string.

\section{Basic formulas of $\overline\partial$-dressing for mKP equation, determinant formula for multi-lump solutions}
\label{Section_2}
\setcounter{equation}{0}
The general formulas of $\overline\partial$-dressing method of Zakharov and Manakov in application to mKP equation are extensively described in~\cite{Konopelchenko&Dubrovsky}, \cite{Konopelchenko&Dubrovsky2}, \cite{KonopelchenkoBook1}, \cite{KonopelchenkoBook2}. Basic object of $\overline\partial$-dressing is a wave function $\chi(\lambda,\overline \lambda; x,y,t)$  which is the function of spectral
variables $\lambda, \overline\lambda$  and spacetime variables $x, y, t$. This function is connected with wave function $\psi(\lambda,\overline\lambda; x,y,t)$ of linear auxiliary problems (\ref{mKP auxiltary problems}) through the formula
\begin{equation}\label{Psi&Chi}
  \psi(\lambda,\overline\lambda; x,y,t):=\chi(\lambda,\overline\lambda; x,y,t)\exp{F(\lambda; x,y,t)},
\end{equation}
here the phase $F(\lambda; x,y,t)$ in exponent is given by expression:
\begin{equation}\label{F(lambda)}
  F(\lambda; x,y,t)=i\frac{x}{\lambda}+\frac{y}{\sigma\lambda^2}+
  \frac{4it}{\lambda^3}
\end{equation}
with $\sigma=i$ - for mKP-1 and $\sigma=1$ - for mKP-2 versions.
Herein instead of $\chi(\lambda,\overline\lambda;x,y,t)$, $F(\lambda; x,y,t)$ etc. we are using the following more short notations: $\chi(\lambda,\overline\lambda)$ , $F(\lambda)$.

Basic equation of $\overline\partial$-dressing method is the so-called $\overline\partial$-problem for wave function $\chi(\lambda,\overline\lambda)$ or equivalent to it singular integral equation for wave function $\chi$:
\begin{equation}\label{di_problem1}
\chi (\lambda,\overline\lambda) = 1 + \frac{2i}{\pi}\iint\limits_\mathbb{C} {\frac{d{\lambda }'_R
d{\lambda'_I}}{\lambda'-\lambda}}
\iint\limits_\mathbb{C}  \chi(\mu,\overline{\mu})
R(\mu ,\overline \mu ;\lambda'
,\overline {\lambda' };x,y,t){d\mu_R  d\mu_I},
\end{equation}
here as usual $\lambda=\lambda_R+i\lambda_I$ and $\mu=\mu_R+i\mu_I$ are complex spectral variables. The kernel $R$ has the form
\begin{equation}
R(\mu,\overline\mu;\lambda,\overline\lambda;x,y,t)=
R_0(\mu,\overline\mu;\lambda,\overline\lambda)e^{F(\mu)-F(\lambda)}.
\end{equation}
Herein we used the canonical normalization $\chi\big|_{\lambda\rightarrow\infty}\rightarrow1$ in (\ref{di_problem1}). Such possibility follows from linear auxiliary problems (\ref{mKP auxiltary problems}) and special dependence on spectral variables $\lambda$  in terms of inverse power $\lambda^{-1}$ in (\ref{F(lambda)}).

Reconstruction formula for $u$
\begin{equation}\label{reconstruct}
u=-\frac{2}{\sigma}\frac{\chi_{0x}}{\chi_0}
\end{equation}
allows to calculate $u$ in terms of zeroth order $\chi_0$ term of Taylor expansion of $\chi(\lambda,\overline\lambda;x,y,t)  =\chi_0(x,y,t)+\lambda\chi_1(x,y,t)+\ldots$ in the neighborhood of $\lambda=0$. From (\ref{di_problem1}) follows the formula for $\chi_0$:
\begin{equation}\label{chi_0}
\chi_0 (x,y,t) = 1 + \frac{2i}{\pi}\iint\limits_\mathbb{C} \frac{d\lambda_R d\lambda_I}{\lambda}
\iint\limits_\mathbb{C}  \chi(\mu,\overline{\mu})R_0(\mu,\overline \mu;\lambda,\overline {\lambda })e^{F(\mu)-F(\lambda)}d\mu_R  d\mu_I.
\end{equation}

For delta-form kernels of the type
\begin{equation}\label{kernel1}
R_0(\mu,\overline{\mu};\lambda,\overline{\lambda})
=\sum\limits_k^{N} A_k\delta(\mu-\mu_k)\delta(\lambda-\lambda_k).
\end{equation}
with complex amplitudes $A_k$ and complex spectral points $\mu_k\neq\lambda_k$, so-called general determinant formula for multi-soliton solutions can be derived~\cite{Konopelchenko&Dubrovsky}, \cite{Konopelchenko&Dubrovsky2}, \cite{KonopelchenkoBook1}, \cite{KonopelchenkoBook2}. Here will be multi-lump solutions considered, for such type of solutions kernel $R_0(\mu,\overline{\mu};\lambda,\overline{\lambda})$ has equal spectral points $\lambda_k=\mu_k$:
\begin{equation}\label{kernel sum}
R_0(\mu,\overline{\mu};\lambda,\overline{\lambda})
=\sum\limits_k^{N} A_k\delta(\mu-\mu_k)\delta(\lambda-\mu_k).
\end{equation}
For given delta-kernel $R_0$ (\ref{kernel sum})
with complex amplitudes $A_k$ and complex \,"spectral"\, points $\mu_k$ one can easily obtain general determinant formula for exact multi-lump  solutions (complex in general) for mKP equation (\ref{mKP}).  We repeated the derivation of well known determinant formula for multi-lump solutions~\cite{KonopelchenkoBook1}, \cite{KonopelchenkoBook2} and give for it convenient form  introducing by the way convenient notations and useful terminology. From (\ref{di_problem1}) and (\ref{kernel sum}) we derived the wave function $\chi(\lambda,\overline\lambda)$
\begin{equation}\label{chi(lambda1)}
\chi(\lambda,\overline\lambda)=1-\frac{2i}{\pi}\sum^N_{k=1}
\frac{A_k}{\lambda-\mu_k}\chi(\mu_k)
\end{equation}
in the form of the sum of $N$ terms with simple poles at \,"spectral"\, points $\mu_k$. Such pole structure of wave function on spectral variables $\lambda$ is typical for quantum mechanics with basic Schr\"{o}dinger equation and corresponding pole structures of quantum-mechanical wave functions from wave numbers, energy, momentum, etc. Formula (\ref{chi(lambda1)}) expresses the wave function  $\chi(\lambda,\overline\lambda)$ at arbitrary values of spectral variables $\lambda, \overline\lambda$ in terms of some kind of "basis" or basic set of $N$ wave functions $\chi(\mu_k):=\chi(\mu_k,\overline{\mu_k})$, $(k=1,\ldots,N)$ at spectral points $\mu_k$ corresponding to the choice (\ref{kernel sum}) of the kernel $R_0$.

From integral equation (\ref{di_problem1}) follows linear algebraic system of equations for the set of wave functions $\chi(\mu_k)\equiv\chi(\mu_k,\overline\mu_k)$, $(k=1,\ldots,N)$:
\begin{equation}\label{tildeA}
\sum\limits_{l=1}^N\tilde{A}_{kl}\chi(\mu_k)=1;\quad \tilde{A}_{kl}=\left(1+\frac{2A_k}{\pi\mu_k}X_k\right)\delta_{kl}+
\frac{2iA_l/\pi}{\mu_k-\mu_l}(1-\delta_{kl}),
\end{equation}
here for convenience we introduced the quantities $X_k=X_k(\mu_k)$  through the formula:
 \begin{equation}\label{dF_formula}
F'(\mu_k):=\frac{\partial F(\mu)}{\partial\mu}\bigg|_{\mu=\mu_k}=-\frac{i}{\mu_k}X_k(\mu_k),\quad X_k=\frac{x}{\mu_k}-\frac{2iy}{\sigma\mu^2_k}+\frac{12t}{\mu_k^3}.
\end{equation}
The details of such kind calculations one can see in ~\cite{KonopelchenkoBook1}, \cite{KonopelchenkoBook2}.
We redefined matrix $\tilde A_{kl}$ in following form:
\begin{equation}\label{matrixA1}
\tilde{A}_{kl}:=A_{kl}\frac{2A_l}{\pi\mu_l},\quad \tilde{A}_{kl}^{-1}:=\frac{\pi\mu_k}{2A_k}A_{kl}^{-1},
\end{equation}
introducing more symmetrical matrix $A_{kl}$:
\begin{equation}\label{MatrixA}
A_{kl}=\left(X_k(\mu_k)+\frac{\pi \mu_k}{2 A_k}\right)\delta_{kl}+\frac{i\mu_l}{\mu_k-\mu_l}(1-\delta_{kl}).
\end{equation}
For the coefficient $\chi_0(x,y,t)$ from (\ref{chi_0}), (\ref{kernel sum}) using  (\ref{tildeA}) and (\ref{MatrixA}) we obtained:
\begin{equation}\label{chi0-1}
\chi_0=1+\sum\limits_{k,l}\frac{2iA_k}{\pi\mu_k}\tilde{A}_{kl}^{-1}=
1+i\sum\limits_{k,l}A_{kl}^{-1}=1+\textrm{tr}(BA^{-1})
=\det(1+BA^{-1}).
\end{equation}
Here matrix $B$
\begin{equation}\label{MatrixB}
B_{kl}=i,\quad \forall\,k,l
\end{equation}
has rank $1$;  matrix $B\cdot A^{-1}$ is also rank $1$. We used in derivation (\ref{chi0-1}) useful matrix identity for rank $1$ matrix $F$
\begin{equation}\label{TrofrankoneMatrix}
1+\textrm{tr} F=\det(1+F).
\end{equation}
Finally reconstruction formula (\ref{reconstruct}) gives
\begin{equation}\label{reconstruct1}
u=-\frac{2}{\sigma}\frac{\partial}{\partial x}\ln\chi_0=-\frac{2}{\sigma}\frac{\partial}{\partial x}\ln\left(\frac{\det(A+B)}{\det A}\right).
\end{equation}
Formula (\ref{reconstruct1}) represents in general determinant form generally complex-valued exact multi-lump solutions of mKP equation with complex parameters $A_k$ and $\mu_k$. 

In order to satisfy boundary condition (\ref{BoundaryCond}) and reality condition $u=\overline u$ for solutions $u$ we had  to specifically choose the complex amplitudes $A_k$ and complex spectral parameters $\mu_k$ of the kernel $R_0(\mu,\overline{\mu};\lambda,\overline{\lambda})$ of $\overline{\partial}$-problem of $\overline{\partial}$-dressing method. In fact the satisfaction of integrable boundary condition (\ref{BoundaryCond}) and of reality condition $u=\overline u$ is central problem in calculations of exact solutions, this requires special consideration and the corresponding calculations are not so simple as it may be seemed.

\section{The restrictions  on parameters of $R_0$ kernel from  integrable boundary condition}
\label{Section_3}
\setcounter{equation}{0}
The restrictions on the kernel $R_0$ of $\overline\partial$-problem from boundary condition (\ref{BoundaryCond}) can be obtained analogously to derivation of restrictions from reality condition \cite{Konopelchenko&Dubrovsky}  by the use of so-called \,"limit of weak fields"\, by the following way.
Due to reconstruction formula (\ref{reconstruct}) or (\ref{reconstruct1})
\begin{equation}\label{rFor}
u(x,y,t)\bigg|_{y=0}=
-\frac{2}{\sigma}\frac{\chi_{0x}}{\chi_{0}}\bigg|_{y=0}=0
\end{equation}
if $\chi_{0x}\big|_{y=0}=0$. We derived from (\ref{chi0-1}) using (\ref{tildeA})-(\ref{MatrixA}):
\begin{eqnarray}
&\chi_{0x}\big|_{y=0}=i\sum\limits_{k,l}\left(A_{kl}^{-1}\right)_x
\big|_{y=0}=-i\sum\limits_{k,l,m,n}A^{-1}_{km}A_{mn,\,x}A^{-1}_{nl}
\big|_{y=0}=\nonumber\\
&-i\sum\limits_{k,l,m}A^{-1}_{km}\frac{1}{\mu_m}A^{-1}_{ml}\big|_{y=0}=
-i\textrm{tr}(GA^{-1}HA^{-1})\big|_{y=0}.
\end{eqnarray}
Here we introduced the matrices $G$ and $H$:
\begin{equation}\label{MatrixBD}
G_{lk}:=1,\,\forall\,k,l;\quad H_{mn}:=\frac{1}{\mu_n}\delta_{mn}=\textrm{diag}\left(\frac{1}{\mu_n}\right).
\end{equation}
 So due to (\ref{TrofrankoneMatrix}) from condition $\chi_{0x}\big|_{y=0}=0$  it follows:
 \begin{eqnarray}
\textrm{tr}(GA^{-1}HA^{-1})\big|_{y=0}=0\Rightarrow \left(1+\textrm{tr}(GA^{-1}HA^{-1})\right)\big|_{y=0}=\nonumber\\
=\left(\frac{\det(A+GA^{-1}H)}{\det A}\right)\bigg|_{y=0}=1,
\end{eqnarray}
i.e. the condition, equivalent to $u\big|_{y=0}=0$, has the form:
 \begin{equation}\label{BoundaryConditionDetForm}
\det(A+GA^{-1}H)\big|_{y=0}=\det A\big|_{y=0},
\end{equation}
and gives the convenient recipe for satisfaction of integrable boundary condition (\ref{BoundaryCond}). We had to choose the parameters $A_k$ and $\mu_k$ and $\lambda_k$ of $R_0$-kernel in such a way that (\ref{BoundaryConditionDetForm}) is satisfied.

Here we demonstrated the effectiveness of using (\ref{BoundaryConditionDetForm}) for the case $N=2$  of  two terms in kernel $R_0(\mu,\overline{\mu};\lambda,\overline{\lambda})$  (\ref{kernel1}) with spectral points $\mu_1$ and $-\mu_1$:
\begin{equation}\label{KR0}
R_0(\mu,\overline{\mu};\lambda,\overline{\lambda})=
A_1\delta(\mu-\mu_1)\delta(\lambda-\mu_1)+A_2\delta(\mu+\mu_1)\delta(\lambda+\mu_1).
\end{equation}
The matrices $A$, $A^{-1}$ due to (\ref{MatrixA}) have the forms:
\begin{equation}\label{matrA}
A:=\left(
\begin{array}{cc}
      \tilde {X}_1 & -\frac{i}{2} \\
      -\frac{i}{2} & \tilde {X}_2\\
\end{array}
\right),
\quad
A^{-1}=\frac{1}{\Delta}\left(
\begin{array}{cc}
      \tilde {X}_2 & \frac{i}{2} \\
      \frac{i}{2} & \tilde {X}_1\\
\end{array}
\right),\quad
\Delta=\tilde {X}_1\tilde {X}_2+\frac{1}{4},
\end{equation}
where
\begin{equation}\label{X1X2}
\tilde {X}_1:=X_1+\frac{\pi\mu_1}{2A_1} \quad
\tilde {X}_2:=X_2-\frac{\pi\mu_1}{2A_2}.
\end{equation}
We obtained by the use of  (\ref{MatrixBD}) and (\ref{matrA}), (\ref{X1X2}):
\begin{equation}\label{A+BAD}
A+GA^{-1}H=
\left(
\begin{array}{cc}
           \tilde{X}_1+\frac{1}{\mu_1\Delta}\left(\tilde{X}_2+\frac{i}{2}\right), & -\frac{1}{\mu_1\Delta}\left(\tilde{X}_1+\frac{i}{2}\right)-\frac{i}{2}\\
            -\frac{i}{2}+\frac{1}{\mu_1\Delta}\left(\tilde{X}_2+\frac{i}{2}\right), &  \tilde{X}_2-\frac{1}{\mu_1\Delta}\left(\tilde{X}_1+\frac{i}{2}\right)  \\
             \end{array}
           \right).
\end{equation}
From (\ref{A+BAD}) and (\ref{BoundaryConditionDetForm}) it follows the relation:
\begin{eqnarray}\label{BoundCond1}
& \det(A+GA^{-1}H)\big|_{y=0}=\left(\tilde{X}_1\tilde{X}_2+\frac{1}{4}+\frac{1}{\Delta \mu_1}\left[\tilde{X}^2_2-\tilde{X}^2_1+
i(\tilde{X}_2-\tilde{X}_1)\right]\right)\big|_{y=0}=\nonumber \\
&\stackrel{(\ref{BoundaryConditionDetForm})}{=}\det A\big|_{y=0}=\tilde{X}_1\tilde{X}_2\big|_{y=0}+\frac{1}{4},
\end{eqnarray}
and finally the integrable boundary condition (\ref{rFor}) due to (\ref{BoundCond1}) is equivalent to the relation:
\begin{equation}\label{BoundaryConditionOnAmlitude}
(\tilde{X}_2-\tilde{X}_1)(\tilde{X}_2+\tilde{X}_1+i)\big|_{y=0}=0.
\end{equation}
Taking into account the definitions (\ref{dF_formula}), (\ref{X1X2}) we derived from (\ref{BoundaryConditionOnAmlitude}) that the first multiplier of last expression
\begin{equation}
(\tilde{X}_2-\tilde{X}_1)\big|_{y=0}=-2\left(\frac{x}{\mu_1}+
\frac{12t}{\mu^3_1}\right)-\frac{\pi\mu_1}{2A_2}-
\frac{\pi\mu_1}{2A_1}\neq0
\end{equation}
is not equal to zero for all $x,t$, but for second constant multiplier we had chosen zero value:
\begin{equation}
(\tilde{X}_2+\tilde{X}_1+i)\big|_{y=0}=
-\frac{\pi\mu_1}{2A_2}+\frac{\pi\mu_1}{2A_1}+i=0.
\end{equation}
So the restriction on parameters $A_1$, $A_2$ and $\mu_1$ of kernel $R_0$  (\ref{KR0}) from boundary condition (\ref{rFor}) has the form:
\begin{equation}\label{BoundaryC}
\frac{\pi\mu_1}{2A_2}-\frac{\pi\mu_1}{2A_1}=i.
\end{equation}
One can show that for more general kernel $R_0$ with paired delta-terms
\begin{equation}
R_0(\mu,\overline\mu;\lambda,\overline\lambda)=\sum\limits^N_{k=1}\left(A_{1k}\delta(\mu-\mu_k)\delta(\lambda-\mu_k)+A_{2k}\delta(\mu+\mu_k)\delta(\lambda+\mu_k)\right)
\end{equation}
similar to (\ref{KR0}) boundary condition (\ref{rFor}) is satisfied if the restrictions on parameters $A_{1k}$, $A_{2k}$, $\mu_k$  of all pairs in the sum have the form:
\begin{equation}\label{BoundaryCk}
\frac{\pi\mu_k}{2A_{2k}}-\frac{\pi\mu_k}{2A_{1k}}=i,\quad (k=1,\ldots,N).
\end{equation}
The results obtained in sections 2, 3 for both versions  mKP-1 ($\sigma=i$) and mKP-2 ($\sigma=1$) are valid, this is true also for important restrictions (\ref{BoundaryCk}) on parameters of kernel $R_0$ from boundary condition (\ref{rFor}).

\section{The restrictions  on parameters of $R_0$ kernel from reality condition for mKP-1 equation}
\label{Section_4}
\setcounter{equation}{0}
The reality condition $u=\overline u$ can be satisfied imposing on the kernel $R_0$  some  restrictions which in the case of mKP equation have the form~\cite{Konopelchenko&Dubrovsky2},~\cite{KonopelchenkoBook2}:
\begin{eqnarray}\label{OldRestr}
\text{mKP-1}:\quad\mu R_0(\mu,\overline\mu;\lambda,\overline\lambda)=
\lambda\overline{R_0(\overline\lambda,\lambda;\overline\mu,\mu)};\nonumber\\
\text{mKP-2}:\quad
R_0(\mu,\overline\mu;\lambda,\overline\lambda)=
\overline{R_0(-\overline\mu,-\mu;-\overline\lambda,-\lambda)}.
\end{eqnarray}
These restrictions  have been obtained in the framework of  $\overline\partial$-dressing method in the so-called \,"limit of weak fields"\,~\cite{Konopelchenko&Dubrovsky2},~\cite{KonopelchenkoBook2}: using approximate value $\chi(\lambda, \overline\lambda)\approx 1$  (as the first iteration for solution of integral equation (\ref{di_problem1})) for the wave function $\chi(\lambda, \overline\lambda)$  in integrand (\ref{chi_0}).
The restrictions (\ref{OldRestr})  on kernels $R_0$  derived in such manner have been used several decades ago
for calculating via $\overline\partial$-dressing method the
 different types of exact real solutions of mKP-1 and mKP-2 equations: solutions with functional parameters, multi-soliton and multi-lump solutions~\cite{Konopelchenko&Dubrovsky2}.

The restrictions (\ref{OldRestr}) from reality condition $\overline u=u$  on the kernel $R_0(\mu,\overline\mu;\lambda,\overline\lambda)$ obtained in the limit of weak fields are not general, their derivation in ~\cite{Konopelchenko&Dubrovsky2}, ~\cite{KonopelchenkoBook2} is not rigorous. In some cases, such as at considered in present paper, these restrictions are not working. The restrictions from reality condition $\overline u=u$ can be obtained directly and rigorously by the use general determinant formula (\ref{reconstruct1}) for exact complex-valued multi-lump solutions. Below it is shown how such approach works for  mKP-1 equation. Derived from reality of $u(x,y,t)$  by direct use of (\ref{reconstruct1}) restrictions on kernel $R_0$ $\overline\partial$-problem are very useful also in calculations of exact solutions with integrable boundaries
and lead to new classes of exact real multi-lump solutions of mKP equation with integrable boundary.

At first for mKP-1 equation with $\sigma=i$ we considered the  kernel to $R_0(\mu,\overline\mu;\lambda,\overline\lambda)$ of the form:
\begin{equation}\label{kernelRealCond}
R_0(\mu,\overline\mu;\lambda,\overline\lambda)=\sum\limits_{k}A_k\delta(\mu-\mu_{k0})
\delta(\lambda-\mu_{k0})
\end{equation}
with complex amplitudes $A_k$ and real spectral parameters $\mu_{k0}=\overline{\mu_{k0}}$.
Due to definitions (\ref{dF_formula}) the quantities
\begin{equation}\label{X RealityCond}
X_k:=i\mu_{k0}F^{'}(\mu_{k0})=\frac{x}{\mu_{k0}}-\frac{2y}{\mu^2_{k0}}+\frac{12t}{\mu^3_{k0}}
=\overline{X}_k,
\end{equation}
are  real functions.
We derived using reconstruction formula (\ref{reconstruct}) and (\ref{reconstruct1}):
\begin{equation}\label{reconstructRealCond}
u=2i\frac{\partial}{\partial x}\ln\frac{\det(A+B)}{\det A}=\overline{u},\quad \text{if}\quad  \det(A+B)=\overline{\det A}.
\end{equation}
Matrices $A^{+}$ (hermitian conjugate to $A$) and $A+B$ for mKP-1, due to (\ref{MatrixA}), (\ref{MatrixB}), (\ref{kernelRealCond}) and (\ref{X RealityCond}),  have  the forms:
\begin{equation}\label{MatrixA+}
(A^{+})_{kl}=\overline{A}_{lk}=\left(X_k+
\frac{\pi\mu_{k0}}{2\overline{A}_k}\right)\delta_{kl}-
\frac{i\mu_{k0}}{\mu_{l0}-\mu_{k0}}(1-\delta_{kl}),
\end{equation}
\begin{equation}\label{MatrixA+B}
(A+B)_{kl}=\left(X_k+\frac{\pi\mu_{k0}}{2A_k}+
i\right)\delta_{kl}+\frac{i\mu_{k0}}{\mu_{k0}-\mu_{l0}}(1-\delta_{kl}).
\end{equation}
In considered case of the kernel $R_0$ (\ref{kernelRealCond}), due to reality $\mu_{k0}$ and $X_k$, one can require in (\ref{MatrixA+}) and (\ref{MatrixA+B}):
\begin{equation}\label{KernelReality}
\frac{\pi\mu_{k0}}{2\overline{A}_k}=\frac{\pi\mu_{k0}}{2A_k}+i,
\end{equation}
then the matrices $A^{+}$ and $A+B$ exactly coincide and condition of reality in the form (\ref{reconstructRealCond}), i.e. $\det(A+B)=\overline{\det A}$, is fulfilled.

As second case we considered pure imaginary points $\mu_k=i\mu_{k0}$ in $R_0$ kernel of the form (\ref{kernel1}). This case requires special attention, indeed in this case the quantities $X_k$, due to their definition (\ref{dF_formula}), have the form:
\begin{equation}
 X_k=\left(\frac{x}{\mu_k}-\frac{2y}{\mu^2_k}+
\frac{12t}{\mu^3_k}\right)\big|_{\mu_k=i\mu_{k0}}=
-i\left(\frac{x}{\mu_{k0}}-
\frac{12t}{\mu^3_{k0}}\right)+\frac{2y}{\mu^2_{k0}}:=
-i\Phi_k(x,t)+\Theta_k(y),
\end{equation}
and $\overline{X}_k=i\Phi_k+\Theta_k\neq X_k$, the trick with (\ref{MatrixA+}) and (\ref{MatrixA+B}) is now not valid. For satisfaction of reality condition (\ref{reconstructRealCond}) we used the paired terms in $R_0$ kernel:
\begin{equation}\label{KernelSumReality}
 R_0(\mu,\overline\mu;\lambda,\overline\lambda)=\sum\limits^N_{k=1}\left(A_{1k}\delta(\mu-i\mu_{k0})
\delta(\lambda-i\mu_{k0})+
A_{2k}\delta(\mu+i\mu_{k0})\delta(\lambda+i\mu_{k0})\right).
\end{equation}
For the kernel (\ref{KernelSumReality}) with one pair of terms ($N=1$) the matrices $\overline{A}$ and $A+B$ have the forms:
\begin{equation}\label{matrixA}
\overline A=
\left(
  \begin{array}{cc}
    i\Phi_1+\Theta_1-\frac{i\mu_{10}\pi}{2\overline A_1} & \frac{i}{2} \\
   \frac{i}{2} &   -i\Phi_1+\Theta_1+\frac{i\mu_{10}\pi}{2\overline A_2}\\
  \end{array}
\right),
\end{equation}
\begin{equation}\label{matrixAB}
 A+B=
\left(
  \begin{array}{cc}
   - i\Phi_1+\Theta_1+\frac{i\mu_{10}\pi}{2 A_1}+i & \frac{i}{2} \\
   \frac{i}{2} &   i\Phi_1+\Theta_1-\frac{i\mu_{10}\pi}{2 A_2}+i \\
  \end{array}
\right),
\end{equation}
the requirement of reality  (\ref{reconstructRealCond}), due to (\ref{matrixA}) and (\ref{matrixAB}), gives in considered case:
\begin{equation}
\frac{\mu_{10}\pi}{2\overline{A}_2}=\frac{\mu_{10}\pi}{2A_1}+1.
\end{equation}
We showed that for general case of $R_0$ kernel (\ref{KernelSumReality}) with $N$ paired terms analogous restrictions
\begin{equation}\label{RealCond2}
\frac{\mu_{k0}\pi}{2\overline{A}_{2k}}=\frac{\mu_{k0}\pi}{2A_{1k}}+1
\end{equation}
for all pairs of terms in (\ref{KernelSumReality}) provide reality of corresponding solutions.

We summarized here the results of sections 3 and 4 concerning the restrictions (\ref{BoundaryCk}) from boundary condition (\ref{BoundaryCond}) and the restrictions (\ref{KernelReality}), (\ref{RealCond2}) from reality condition $u=\overline u$  for mKP-1 equation with $\sigma=i$. We established (justified) two possible choices of the kernel $R_0$ with paired terms.

\textbf{A. } For the kernel $R_0$ with pure real spectral points:
\begin{equation}\label{KernelVarA}
 R_0(\mu,\overline\mu;\lambda,\overline\lambda)=\sum\limits^N_{k=1}\left(A_{1k}\delta(\mu-\mu_{k0})\delta(\lambda-\mu_{k0})+
A_{2k}\delta(\mu+\mu_{k0})\delta(\lambda+\mu_{k0})\right),
\end{equation}
where $\overline{\mu}_{k0}=\mu_{k0}$, the restrictions on parameters $A_{1k}$, $A_{2k}$ and $\mu_{k0}$ ($k=1,\ldots,N$) from boundary and reality conditions due to (\ref{BoundaryCk}) and (\ref{KernelReality}) are the following:
\begin{equation}\label{BoundaryCondVarA}
\frac{\pi\mu_{k0}}{2A_{2k}}-\frac{\pi\mu_{k0}}{2A_{1k}}=i -  \text{boundary},
\end{equation}
\begin{equation}\label{RealityCondVarA}
\frac{\pi\mu_{k0}}{2\overline{A}_{1k}}=\frac{\pi\mu_{k0}}{2A_{1k}}+i,\quad \frac{\pi\mu_{k0}}{2\overline{A}_{2k}}=\frac{\pi\mu_{k0}}{2A_{2k}}-i -\text{reality}.
\end{equation}

\textbf{B. } For the kernel $R_0$ with pure imaginary spectral points:
\begin{equation}\label{KernelVarB}
R_0(\mu,\overline\mu;\lambda,\overline\lambda)=\sum\limits^N_{k=1}\left(A_{1k}\delta(\mu-i\mu_{k0})\delta(\lambda-i\mu_{k0})+
A_{2k}\delta(\mu+i\mu_{k0})\delta(\lambda+i\mu_{k0})\right),
\end{equation}
where $\overline{\mu}_{k0}=\mu_{k0}$, the restrictions on parameters $A_{1k}$, $A_{2k}$ and $\mu_{k0}$ ($k=1,\ldots,N$) from boundary and reality conditions, due to (\ref{BoundaryCk}) and (\ref{RealCond2}), are the following:
\begin{equation}\label{BoundaryCondVarB}
\frac{\pi\mu_{k0}}{2A_{2k}}-\frac{\pi\mu_{k0}}{2A_{1k}}=1 -  \text{boundary},
\end{equation}
\begin{equation}\label{RealityCondVarB}
\frac{\pi\mu_{k0}}{2\overline{A}_{2k}}=\frac{\pi\mu_{k0}}{2A_{1k}}+1 - \text{reality}.
\end{equation}

In the next section the simplest exact real two-lump solutions of mKP-1 equation with kernels $R_0$ (\ref{KernelVarA}) and (\ref{KernelVarB})  calculated  by the use of general determinant formula (\ref{reconstructRealCond}) are presented.

\section{Two-lump solutions of mKP-1 equation for the kernel $R_0$ with real and imaginary spectral points}
\label{Section_5}
\begin{figure}[h]
\begin{center}
\includegraphics[width=0.50\textwidth,keepaspectratio]{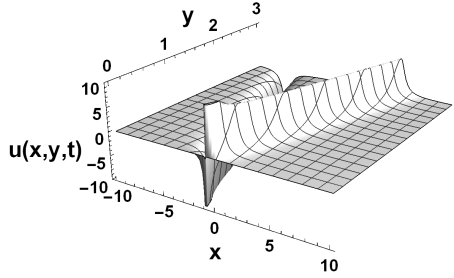}
\parbox[t]{1\textwidth}{\caption{Lump solution $u$ (\ref{Solution mKP1 1})  with parameter $\mu_{10}=1$, $\gamma_1=1$.}\label{Lump(RealSpectral)mKP1}}
\end{center}
\end{figure}

\setcounter{equation}{0}
For the simplest kernel (\ref{KernelVarA}) with one paired terms
\begin{equation}\label{KernelVarAexample}
R_0(\mu,\overline\mu;\lambda,\overline\lambda)=A_{1}\delta(\mu-\mu_{10})\delta(\lambda-\mu_{10})+
A_{2}\delta(\mu+\mu_{10})\delta(\lambda+\mu_{10}),
\end{equation}
we had from general formulas (\ref{X RealityCond}):
\begin{align}\label{X RealityCondExample1}
X_1(\mu_{10})=\frac{x}{\mu_{10}}-&\frac{2y}{\mu^2_{10}}+\frac{12t}{\mu^3_{10}}:
=\tilde{\Phi}(x,t)-\Theta(y),\quad X_2(-\mu_{10})=-\tilde{\Phi}(x,t)-\Theta(y),\nonumber \\
&\tilde{\Phi}(x,t)=\frac{x}{\mu_{10}}+\frac{12t}{\mu^3_{10}},\quad
\Theta(y)=\frac{2y}{\mu^2_{10}}.
\end{align}
Due to (\ref{BoundaryCondVarA}), (\ref{RealityCondVarA}) the following convenient parametrization of parameters $A_1$, $A_2$, $\mu_{10}$ is valid:
\begin{equation}\label{parametrisation_mKP1}
\frac{\pi\mu_{10}}{2A_1}=\gamma_1-\frac{i}{2},\quad \frac{\pi\mu_{10}}{2A_2}=\gamma_2+\frac{i}{2}=\gamma_1+\frac{i}{2},
\end{equation}
where $\gamma_1$, $\gamma_2$ - are real constants.
The matrix $A$ due to  (\ref{MatrixA}) and (\ref{X RealityCondExample1}),  (\ref{parametrisation_mKP1}) has the form:
\begin{equation}
 A=
\left(
  \begin{array}{cc}
    \tilde\Phi-\Theta+\gamma_1-\frac{i}{2} & -\frac{i}{2} \\
-\frac{i}{2}& -\tilde\Phi-\Theta-\gamma_1-\frac{i}{2} \\
  \end{array}
\right):=
\left(
  \begin{array}{cc}
    \Phi-\Theta-\frac{i}{2} & -\frac{i}{2} \\
    -\frac{i}{2} & -\Phi-\Theta-\frac{i}{2}\\
  \end{array}
\right),
\end{equation}
where $\Phi:=\tilde\Phi+\gamma_1=\frac{x}{\mu_{10}}+\frac{12t}{\mu^3_{10}}+
\gamma_1$.
We obtained for determinant of matrix $A$:
\begin{equation}
\det A=\Theta^2-\Phi^2+i\Theta.
\end{equation}
Then according to reconstruction formula (\ref{reconstructRealCond}) and to reality condition $\det(A+B)=\overline{\det A}$ we obtained:
\begin{equation}\label{Solution mKP1 1}
 u=2i\frac{\partial}{\partial x}\left(-2i\arg\det A\right)=4\frac{\partial}{\partial x}\arctan\frac{\Theta}{\Theta^2-\Phi^2}=
\frac{(8/\mu_{10})\Theta(y)\Phi(x,t)}{(\Theta^2-\Phi^2)^2+\Theta^2}.
\end{equation}
This exact two-lump localized solution   (see figure (\ref{Lump(RealSpectral)mKP1})) of mKP-1 equation with integrable boundary (\ref{BoundaryCond}) has point singularities and represents \,"bound state"\, of two localized single lumps moving together along $x$-axis with velocity $V_x=-\frac{12}{\mu^2_{10}}$. This is certain eigenmode of the field $u(x,y,t)$ in semi-plane $y\geq 0$ arising due to imposition of boundary condition (\ref{BoundaryCond}).

\begin{figure}[h]
\begin{center}
\includegraphics[width=0.50\textwidth,keepaspectratio]{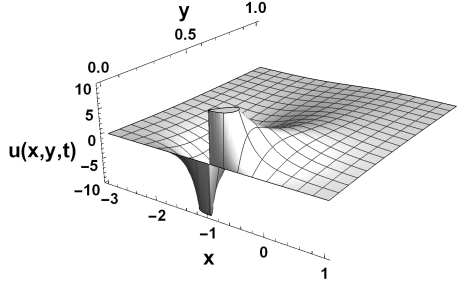}
\parbox[t]{1\textwidth}{\caption{Lump solution $u$ (\ref{Solution mKP1 2}) with parameter $\mu_{10}=1$, $\gamma_1=1$.}\label{Lump(ImaginarySpectral)mKP1}}
\end{center}
\end{figure}

For the simplest kernel  (\ref{KernelSumReality}) with one paired terms with imaginary spectral points:
\begin{equation}\label{KernelVarBexample}
R_0(\mu,\overline\mu;\lambda,\overline\lambda)=A_{1}\delta(\mu-i\mu_{10})\delta(\lambda-i\mu_{10})+
A_{2}\delta(\mu+i\mu_{10})\delta(\lambda+i\mu_{10}),
\end{equation}
we had  from general formulas (\ref{X RealityCond}) and (\ref{KernelVarBexample}):
\begin{align}\label{X RealityCondExample2}
 X_1(i\mu_{10})=-i\left(\frac{x}{\mu_{10}}-
\frac{12t}{\mu_{10}^3}\right)+&\frac{2y}{\mu_{10}^2}:=
-i\tilde{\Phi}(x,t)+\Theta(y),\quad  X_2(-i\mu_{10})=i\tilde{\Phi}(x,t)+\Theta(y),\nonumber \\
&\tilde{\Phi}(x,t):=\frac{x}{\mu_{10}}-\frac{12t}{\mu^3_{10}},\quad \Theta(y):=\frac{2y}{\mu^2_{10}}.
\end{align}
The matrix $A$ due to (\ref{MatrixA}), (\ref{BoundaryCondVarB}), (\ref{RealityCondVarB}) and (\ref{X RealityCondExample2}) has the form:
\begin{equation}\label{MatrixAmKP1ex2}
 A=
\left(
  \begin{array}{cc}
    -i\tilde\Phi+\Theta+i\left(\gamma_1-\frac{1}{2}\right)& -\frac{i}{2} \\
-\frac{i}{2}&  i\tilde\Phi+\Theta-i\left(\gamma_1+\frac{1}{2}\right) \\
  \end{array}
\right):= \left(\begin{array}{cc}
-i\Phi+\Theta-\frac{i}{2}& -\frac{i}{2} \\
-\frac{i}{2}&  i\Phi+\Theta-\frac{i}{2} \\
  \end{array}\right),
\end{equation}
here $\Phi: =\tilde\Phi-\gamma_1=\frac{x}{\mu_{10}}-\frac{12t}{\mu^3_{10}}-
\gamma_1$ and
 via relation (\ref{RealityCondVarB}) we introduced convenient parametrization
\begin{equation}
\frac{\pi\mu_{10}}{2A_1}=\gamma_1-\frac{1}{2},\quad \frac{\pi\mu_{10}}{2A_2}=\gamma_1+\frac{1}{2}
\end{equation}
 with real constant $\overline{\gamma}_1=\gamma_1$.
From (\ref{MatrixAmKP1ex2}) we  derived the expression for $\det A$:
\begin{equation}
\det A=\Theta^2+\Phi^2-i\Theta,
\end{equation}
finally by the reconstruction formula (\ref{reconstructRealCond}) we obtained the exact two-lump solution of mKP-1 equation corresponding to the kernel (\ref{KernelVarBexample}):
\begin{equation}\label{Solution mKP1 2}
u=4\frac{\partial}{\partial x}\arg\det A=-4\frac{\partial}{\partial x}\arctan\frac{\Theta}{\Theta^2+\Phi^2}=\frac{(8/\mu_{10})\Theta(y)
\Phi(x,t)}{(\Theta^2+\Phi^2)^2+\Theta^2}.
\end{equation}
The graph of this solution is shown on figure (\ref{Lump(ImaginarySpectral)mKP1}).
This exact two-lump localized solution  of mKP-1 equation with integrable boundary (\ref{BoundaryCond}) has  point singularities and represents \,"bound state"\, of two localized single lumps moving together along $x$-axis  with velocity $V_x=\frac{12}{\mu^2_{10}}$, this is certain eigenmode of the field $u(x,y,t)$ in semi-plane $y\geq 0$ arising due to imposition of boundary condition (\ref{BoundaryCond}).

\section{The restrictions  on parameters of $R_0$ kernel from reality condition for mKP-2 equation}
\label{Section_6}
\setcounter{equation}{0}
Basic formulas of $\overline\partial$-dressing method for mKP-2 equation are the same as for mKP-1 with the change $\sigma=i$ on $\sigma=1$ in corresponding places. So in present case all formulas from section 2 are valid.
In analogy with mKP-1 case described in section 4 we obtained the restrictions on parameters of $R_0$ which lead to fulfillment of reality condition $u(x,y,t)=\overline{u(x,y,t)}$ .

For the delta-form kernel $R_0$ (\ref{kernel sum}) with pure imaginary spectral parameters $\mu_k=i\mu_{k0}$, $\mu_{k0}=\overline\mu_{k0}$
\begin{equation}\label{kernmKP2}
R_0(\mu,\overline\mu;\lambda,\overline\lambda)=
\sum_{k}A_{k}\delta(\mu-i\mu_{k0})\delta(\lambda-i\mu_{k0})
\end{equation}
 we derived from (\ref{dF_formula}):
\begin{equation}\label{xImaginMu}
X(i\mu_{k0})=-i\left(\frac{x}{\mu_{k0}}-\frac{2y}{\mu^2_{k0}}-
\frac{12t}{\mu^3_{k0}}\right).
\end{equation}
Consequently the quantities $X(i\mu_{k0})$ are pure imaginary: $\overline X(i\mu_{k0})=-X(i\mu_{k0})$. We redefined for convenience matrices $A$ and $A+B$ (\ref{MatrixA}), (\ref{MatrixB}) factoring imaginary unit $-i$:
\begin{eqnarray}\label{newMatrixA&B}
 A_{kl}=-i\left[\left(X_k-\frac{\pi\mu_{k0}}{2A_{k}}\right)
\delta_{kl}-\frac{\mu_{l0}}{\mu_{k0}-\mu_{l0}}(1-\delta_{kl})\right]:=-i D_{kl}, \nonumber \\
 (A + B)_{kl}=-i\left[\left(X_k-\frac{\pi\mu_{k0}}{2A_{k}}-1\right)\delta_{kl}-
\frac{\mu_{k0}}{\mu_{k0}-\mu_{l0}}(1-\delta_{kl})\right]:=-iN_{kl}.
\end{eqnarray}
From reconstruction formula (\ref{reconstruct1}) we obtained general determinant formula for exact solutions of mKP-2 in terms of matrices $N$ and $D$:
\begin{equation}\label{recFormula_mKP2}
u=-2\frac{\partial}{\partial x}\ln\frac{\det N}{\det D}.
\end{equation}
These generally complex-valued solutions, due to definitions (\ref{newMatrixA&B}), will be real for $\det N=\overline {\det N}$, $ \det D= \overline {\det D}$ if the following restrictions  on parameters $A_k, \mu_k, (k=1,...,N)$ of the kernel (\ref{kernmKP2}) are valid:
\begin{equation}\label{reality_imaginary_mKP2}
A_k=\overline{A}_k \quad \mu_k=i\mu_{k0},\quad \mu_{k0}=\overline\mu_{k0}.
\end{equation}

For the kernel $R_0$ (\ref{kernel sum}) with real spectral parameters $\mu_k=\mu_{k0},\quad \mu_{k0}=\overline\mu_{k0}$  the quantities $X_k$, due to their definition (\ref{dF_formula}), have the form:
\begin{equation}
X_k(\mu_{k0}):=\frac{x}{\mu_{k0}}+\frac{12t}{\mu^3_{k0}}-
i\frac{2y}{\mu^2_{k0}}:=\Phi_k(x,t)-i\Theta_k(y),
\end{equation}
with
\begin{equation}\label{PhasesmKP2_1}
\Phi_k(x,t):=\frac{x}{\mu_{k0}}+\frac{12t}{\mu^3_{k0}},\quad \Theta_k(y):=\frac{2y}{\mu^2_{k0}}.
\end{equation}
For satisfaction of reality condition $u=\overline u$ in this case we used the paired terms in $R_0$ kernel:
\begin{equation}\label{KernelSumReality_mKP2}
R_0(\mu,\overline\mu;\lambda,\overline\lambda)=\sum\limits^N_{k=1}\left(A_{1k}\delta(\mu-\mu_{k0})
\delta(\lambda-\mu_{k0})+
A_{2k}\delta(\mu+\mu_{k0})\delta(\lambda+\mu_{k0})\right).
\end{equation}
For the kernel (\ref{KernelSumReality_mKP2}) with one pair of terms ($N=1$) the matrices $A$ and $A+B$ (\ref{MatrixA}), (\ref{MatrixB}) have the forms:
\begin{eqnarray}\label{Matrix(A+B)mKP2}
A=
\left(
  \begin{array}{cc}
    \Phi-i\Theta+\frac{\pi\mu_{10}}{2A_1} & -\frac{i}{2} \\
    -\frac{i}{2} & -\Phi-i\Theta-\frac{\pi\mu_{10}}{2A_2} \\
  \end{array}
\right),\nonumber\\
A+B=
\left(
  \begin{array}{cc}
    \Phi-i\Theta+\frac{\pi\mu_{10}}{2A_1}+i & \frac{i}{2} \\
    \frac{i}{2} & -\Phi-i\Theta-\frac{\pi\mu_{10}}{2A_2}+i \\
  \end{array}
\right).
\end{eqnarray}
From reconstruction formula for mKP-2 (\ref{reconstruct1}) ($\sigma=1$) it follows that conditions of reality $u=\overline u$ is fulfilled if
\begin{equation}\label{mKP2RealCond}
\det A=\overline{\det A},\quad \det(A+B)=\overline{\det(A+B)}.
\end{equation}
We concluded from expression of $A$ (\ref{Matrix(A+B)mKP2}):
\begin{equation}
-\det A=
\Theta^2+\Phi^2-\frac{1}{4}+\frac{\pi\mu_{10}}{2A_2}\Phi+
\frac{\pi\mu_{10}}{2A_1}\Phi-i\Theta\frac{\pi\mu_{10}}{2A_2}+
i\Theta\frac{\pi\mu_{10}}{2A_1}+\frac{\pi^2\mu^2_{10}}{4A_1A_2}.
\end{equation}
The requirement $\det A=\overline{\det A}$ gives finally the relation between amplitudes $A_1$ and $A_2$:
\begin{equation}\label{realityCond_mKP2(1)}
A_2=\overline{A}_1.
\end{equation}
So taking into account (\ref{realityCond_mKP2(1)}) we derived for $-\det A$:
\begin{equation}\label{det(A)mKP2}
-\det A=\Theta^2+\Phi^2+\frac{\pi\mu_{10}A_{1R}}{|A_1|^2}\Phi+
\frac{\pi\mu_{10}A_{1I}}{|A_1|^2}\Theta+\frac{\pi^2\mu^2_{10}}{4|A_1|^2}-
\frac{1}{4}.
\end{equation}

The corresponding calculations with exact expression for $A+B$, due to definition in (\ref{Matrix(A+B)mKP2}), we made quite analogously:
\begin{eqnarray}\label{det(A+B)mKP2}
-\det(A+B)=\Theta^2+\Phi^2-2\Theta+\frac{3}{4}+
\left(\frac{\pi\mu_{10}}{2A_1}+\frac{\pi\mu_{10}}{2A_2}\right)\Phi+
\nonumber \\
+\left(\frac{i\pi\mu_{10}}{2A_1}-
\frac{i\pi\mu_{10}}{2A_2}\right)\Theta+
\frac{i}{2}\pi\mu_{10}\left(\frac{1}{A_2}-\frac{1}{A_1}\right)+
\frac{\pi^2\mu^2_{10}}{4A_1A_2}.
\end{eqnarray}
From (\ref{mKP2RealCond}), (\ref{det(A+B)mKP2}) follows the same restrictions $A_2=\overline{A}_1$ as in (\ref{realityCond_mKP2(1)}). So taking into account (\ref{realityCond_mKP2(1)}) for $-\det(A+B)$ we obtained:
\begin{equation}\label{det(A+B)mKP2(2)}
-\det(A+B)=\Theta^2+\Phi^2-2\Theta+\frac{3}{4}+\frac{\pi\mu_{10}A_{1R}}{|A_1|^2}\Phi+\frac{\pi\mu_{10}A_{1I}}{|A_1|^2}\Theta-\frac{\pi\mu_{10}A_{1I}}{|A_1|^2}+
\frac{\pi^2\mu^2_{10}}{4|A_1|^2}.
\end{equation}
We showed that for general case of $R_0$ kernel with $N$ paired terms analogous restrictions
\begin{equation}\label{RealCond2_mKP2}
A_{2k}=\overline{A}_{1k} \quad \mu_k=\mu_{k0}=\overline\mu_{k0}.
\end{equation}
for all pairs of terms in (\ref{KernelSumReality_mKP2}) provide reality of corresponding solutions.

We summarized here the results of sections 3 and 6 concerning  the restriction (\ref{BoundaryCk}) from boundary condition (\ref{BoundaryCond}) and the restrictions (\ref{reality_imaginary_mKP2}), (\ref{RealCond2_mKP2}) from reality condition $u=\overline u$  for mKP-2 equation with $\sigma=1$. We established (justified) two possible choices of kernel $R_0$ with paired terms.

\textbf{A. } For the kernel $R_0$ with pure real spectral points:
\begin{equation}\label{KernelVarA_mKP2}
 R_0(\mu,\overline\mu;\lambda,\overline\lambda)=\sum\limits^N_{k=1}\left(A_{1k}\delta(\mu-\mu_{k0})\delta(\lambda-\mu_{k0})+
A_{2k}\delta(\mu+\mu_{k0})\delta(\lambda+\mu_{k0})\right),
\end{equation}
where $\overline{\mu}_{k0}=\mu_{k0}$, the restrictions on parameters $A_{1k}$, $A_{2k}$ and $\mu_{k0}$ ($k=1,\ldots,N$) from boundary and reality conditions due to (\ref{BoundaryCk}) and (\ref{RealCond2_mKP2}) are the following:
\begin{equation}\label{BoundaryCondVarA_mKP2}
\frac{\pi\mu_{k0}}{2A_{2k}}-\frac{\pi\mu_{k0}}{2A_{1k}}=i - \text{boundary},
\end{equation}
\begin{equation}\label{RealityCondVarA_mKP2}
A_{2k}=\overline{A}_{1k} - \text{reality}.
\end{equation}

\textbf{B. } For the kernel $R_0$ with pure imaginary spectral points:
\begin{equation}\label{KernelVarB_mKP2}
 R_0(\mu,\overline\mu;\lambda,\overline\lambda)=\sum\limits^N_{k=1}\left(A_{1k}\delta(\mu-i\mu_{k0})\delta(\lambda-i\mu_{k0})+
A_{2k}\delta(\mu+i\mu_{k0})\delta(\lambda+i\mu_{k0})\right),
\end{equation}
where $\overline{\mu}_{k0}=\mu_{k0}$, the restrictions on parameters $A_{1k}$, $A_{2k}$ and $\mu_{k0}$ ($k=1,\ldots,N$) from boundary and reality conditions, due to (\ref{BoundaryCk}) and (\ref{reality_imaginary_mKP2}), have the form:
\begin{equation}\label{BoundaryCondVarB_mKP2}
\frac{\pi\mu_{k0}}{2A_{2k}}-\frac{\pi\mu_{k0}}{2A_{1k}}=1 -  \text{boundary},
\end{equation}
\begin{equation}\label{RealityCondVarB_mKP2}
A_k=\overline{A}_k - \text{reality}.
\end{equation}

In the next section the simplest exact real two-lump solutions of mKP-2 equation with kernels $R_0$ (\ref{KernelVarA_mKP2}) and (\ref{KernelVarB_mKP2}) calculated by the use of general determinant formula (\ref{reconstructRealCond}) are presented.

\section{Two-lump solutions of mKP-2 equation for the kernel $R_0$ with real and imaginary spectral points}
\label{Section_7}
\begin{figure}[h]
\begin{center}
\includegraphics[width=0.50\textwidth,keepaspectratio]{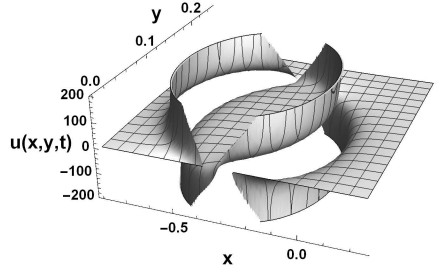}
\parbox[t]{1\textwidth}{\caption{Lump solution $u$ (\ref{lump_mKP2(1)})  with parameter $A_{1R}=A_{1I}=1$.}\label{Lump(RealSpectral)mKP2}}
\end{center}
\end{figure}

\setcounter{equation}{0}
As the first example we calculated exact real two-lump solution of mKP-2 corresponding to the simplest delta-form kernel $R_0$ with real spectral parameters $\mu_{10}$ and $(-\mu_{10})$ of the form (\ref{KernelVarA_mKP2}):
\begin{equation}\label{kernel_mKP2}
R_0(\mu,\overline\mu;\lambda,\overline\lambda)=
A_1\delta(\mu-\mu_{10})\delta(\lambda-\mu_{10})+
A_2\delta(\mu+\mu_{10})\delta(\lambda+\mu_{10}).
\end{equation}
The restrictions (\ref{BoundaryCondVarA_mKP2}) from integrable boundary condition $u\big|_{y=0}=0$ and from reality condition (\ref{RealityCondVarA_mKP2}), taking into account together, lead to relation:
\begin{equation}\label{Boundary&Real}
\frac{\pi\mu_{10}A_{1I}}{|A_1|^{2}}=1.
\end{equation}
By the use of (\ref{Boundary&Real}) we obtained after simplifications the following simple expressions for $\det A$ (\ref{det(A)mKP2}) and $\det(A+B)$ (\ref{det(A+B)mKP2(2)}):
\begin{eqnarray}
-\det A=\left(\Theta+\frac{1}{2}\right)^2+\left(\Phi+
\frac{A_{1R}}{2A_{1I}}\right)^2-\frac{1}{4}:=D_1 , \nonumber \\
-\det (A+B)=\left(\Theta-\frac{1}{2}\right)^2+\left(\Phi+
\frac{A_{1R}}{2A_{1I}}\right)^2-\frac{1}{4}:=D_2.
\end{eqnarray}
Using reconstruction formula (\ref{reconstruct1}) with $\sigma=1$ we calculated the exact two-lump solution of mKP-2 equation with integrable boundary (\ref{BoundaryCond}) corresponding to the kernel $R_0$ (\ref{kernel_mKP2}):
\begin{eqnarray}\label{lump_mKP2(1)}
 u=-2\frac{\partial}{\partial x}\ln\frac{\det(A+B)}{\det A}=-\frac{4\left(\Phi+\frac{A_{1R}}{2A_{1I}}\right)\frac{1}{\mu_{10}}}{D_2}+
\frac{4\left(\Phi+\frac{A_{1R}}{2A_{1I}}\right)\frac{1}{\mu_{10}}}{D_1}=
\nonumber \\
=-8\frac{\Theta(y)\frac{1}{\mu_{10}}\left(\Phi+\frac{A_{1R}}{2A_{1I}}
\right)}{\left[\left(\Theta+\frac{1}{2}\right)^2+\left(\Phi+
\frac{A_{1R}}{2A_{1I}}\right)^2-\frac{1}{4}\right]
\left[\left(\Theta-\frac{1}{2}\right)^2+\left(\Phi+
\frac{A_{1R}}{2A_{1I}}\right)^2-\frac{1}{4}\right]},
\end{eqnarray}
here the phases $\Phi(x,t)$ and $\Theta(y)$ are given by expressions (\ref{PhasesmKP2_1}):
\begin{equation}\label{PhasesmKP2_11}
\Phi(x,t):=\frac{x}{\mu_{10}}+\frac{12t}{\mu^3_{10}},\quad \Theta(y):=\frac{2y}{\mu^2_{10}}.
\end{equation}
Evidently this exact localized two-lump solutions has singularities located
 on two separate circles of equal radii.
The graph of (\ref{lump_mKP2(1)}), shown on figure (\ref{Lump(RealSpectral)mKP2}) at semiplane $y\geq 0$. The expression (\ref{lump_mKP2(1)}) represents two coherently bounded with each other simple lumps (two lumps in \,"bound state") moving with the same velocity $V_x=-\frac{12}{\mu^2_{10}}$ along axes $x$.

As second interesting example we calculated new exact real
two-lump solutions of mKP-2 equation corresponding to the kernel $R_0(\mu,\overline\mu;\lambda,\overline\lambda)$ (\ref{KernelVarB_mKP2}) with imaginary spectral points $i\mu_{10}$ and $(-i\mu_{10})$:
\begin{equation}\label{kernel_mKP2_2}
 R_0(\mu,\overline\mu;\lambda,\overline\lambda)=
A_1\delta(\mu-i\mu_{10})\delta(\lambda-i\mu_{10})+
A_2\delta(\mu+i\mu_{10})\delta(\lambda+i\mu_{10}).
\end{equation}
The restriction (\ref{BoundaryCondVarB_mKP2}) from the boundary condition $u\big|_{y=0}=0$ has form:
\begin{equation}\label{BoundaryConditionMKP2_2}
\frac{\mu_{10}\pi}{2A_2}-\frac{\mu_{10}\pi}{2A_1}=1.
\end{equation}
From (\ref{BoundaryConditionMKP2_2}) follows convenient parametrization:
\begin{equation}\label{Parametrization_mKP2_2}
\frac{\mu_{10}\pi}{2A_1}=-\gamma_1-\frac{1}{2},\quad \frac{\mu_{10}\pi}{2A_2}=-\gamma_1+\frac{1}{2}.
\end{equation}
From the restriction (\ref{RealityCondVarB_mKP2}) of reality condition $u=\overline u$ we concluded that amplitudes
$A_1$ and $A_2$ are real:
\begin{equation}\label{realityCond_mKP2(1)2}
\overline{A}_1=A_1,\quad \overline{A}_2=A_2.
\end{equation}
The quantities $X(i\mu_{10})$ and $X(-i\mu_{10})$ due to (\ref{xImaginMu}) have the forms:
\begin{eqnarray}
&X(i\mu_{10}):=-i(\Theta(y)+\tilde\Phi(x,t)),\quad X(-i\mu_{10}):=-i(\Theta(y)-\tilde\Phi(x,t)),\nonumber \\
&\Theta(y):=-\frac{2y}{\mu^2_{10}},\quad \tilde\Phi(x,t)=\frac{x}{\mu_{10}}-\frac{12t}{\mu^3_{10}}.
\end{eqnarray}
Matrices $A$ and $A+B$ in considered case due to (\ref{xImaginMu}) are given by expressions:
\begin{equation}\label{Matrix(A)mKP2_2}
 A=
\left(
  \begin{array}{cc}
    \Phi+\Theta+\frac{1}{2} & \frac{1}{2} \\
    \frac{1}{2} & -\Phi+\Theta+\frac{1}{2} \\
  \end{array}
\right),\quad
A+B=
\left(
  \begin{array}{cc}
     \Phi+\Theta-\frac{1}{2} & -\frac{1}{2} \\
    -\frac{1}{2} & -\Phi+\Theta-\frac{1}{2} \\
  \end{array}
\right),
\end{equation}
here phase $\Phi(x,t)$ has the form:
\begin{equation}
\Phi(x,t):=\tilde\Phi+\gamma_1=\frac{x}{\mu_{10}}+
\frac{12t}{\mu^3_{10}}+\gamma_1.
\end{equation}

\begin{figure}[h]
\begin{center}
\includegraphics[width=0.50\textwidth,keepaspectratio]{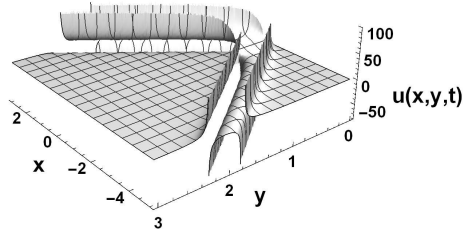}
\parbox[t]{1\textwidth}{\caption{Lump solution $u$ (\ref{Reconstruction_mKP2(2)})  with parameters $\mu_{10}=1$, $\gamma_1=1$.}\label{Lump(ImaginarySpectral)mKP2}}
\end{center}
\end{figure}

From (\ref{Matrix(A)mKP2_2})  we obtained  for the determinants of $A$ and $A+B$:
\begin{equation}\label{D1D2mKP2}
 D_1:=\det A=\left(\frac{1}{2}+\Theta\right)^2-\Phi^2-\frac{1}{4},\quad
D_2:=\det (A+B)=\left(\frac{1}{2}-\Theta\right)^2-\Phi^2-\frac{1}{4}.
\end{equation}
Via reconstruction formula (\ref{reconstruct1}) and (\ref{D1D2mKP2}) we calculated  exact real two-lump solution of mKP-2 equation with integrable boundary (\ref{BoundaryCond}):
\begin{eqnarray}\label{Reconstruction_mKP2(2)}
 u=-2\frac{\partial}{\partial x}\ln\frac{\det(A+B)}{\det A}=-\frac{4\Phi}{\mu_{10}}\left(\frac{1}{D_1}-\frac{1}{D_2}\right)=\nonumber \\
\frac{(8/\mu_{10})\Phi\Theta}{\left[\left(\Theta+\frac{1}{2}\right)^2-\Phi^2-\frac{1}{4}\right]
\left[\left(\Theta-\frac{1}{2}\right)^2-\Phi^2-\frac{1}{4}\right]}.
\end{eqnarray}
Evidently this exact two-lump solution has singularities located at separate hyperbolas. The graph of this solution  is shown on
figure  (\ref{Lump(ImaginarySpectral)mKP2}).
The expression (\ref{Reconstruction_mKP2(2)}) represents two coherently bounded with each other simple lumps (two lumps in\,"bound state") moving with the same velocity $V_x=\frac{12}{\mu^2_{10}}$ along axes $x$.

\section{Conclusions}
\label{Section_8}
\setcounter{equation}{0}
In the present paper we described new classes of exact multi-lump solutions of mKP-1,2 equations with integrable boundary condition $u(x,y,t)\big|_{y=0}=0$. We developed general scheme for calculations of such solutions in framework of $\overline\partial$-dressing method. We showed, that reality condition for solutions $u(x,y,t)$ can be effectively satisfied exactly, imposing the requirement of reality $u=\overline u$ on corresponding exact complex-valued solutions in determinant form; this leads to some restrictions on parameters of solutions, i.e. on amplitudes $A_k$ and spectral points $\mu_k$, $\lambda_k$ of delta-form kernel $R_0$  of $\overline\partial$-problem.

We presented the simplest explicit examples of two-lump solutions of mKP-1,2 equations with integrable boundary: the fulfillment of boundary condition is achieved via special nonlinear superpositions   of two more simpler one-lump solutions, certain
eigenmodes of the field $u(x,y,t)$ in semiplane $y\geq 0$ as analogs of standing wave on string.

We demonstrated effectiveness of $\overline\partial$-dressing method in calculations of multi-lump solutions of several types for mKP-1,2 equations with integrable boundary condition. The developed for this in present paper procedure for calculation via $\overline\partial$-dressing of new classes of exact real multi-lump solutions of mKP equation can be effectively applied to all other integrable (2+1)-dimensional nonlinear equations, some of these studies are currently in progress and the results will be published elsewhere.


\end{document}